\documentclass[longauth,traditabstract]{aa}

\usepackage{graphicx}
\usepackage{longtable}

\usepackage{hyperref}
\usepackage{natbib}                 
\bibpunct{(}{)}{;}{a}{}{,}          
\usepackage{amsmath}                

\usepackage{txfonts}

\begin{document}

   \title{Simultaneous multi-wavelength campaign on PKS~2005-489 in a high state}


 {\small  
\author{The H.E.S.S. Collaboration\\
A.~Abramowski \inst{1}
\and F.~Acero \inst{2}
\and F.~Aharonian \inst{3,4,5}
\and A.G.~Akhperjanian \inst{6,5}
\and G.~Anton \inst{7}
\and A.~Barnacka \inst{8,9}
\and U.~Barres~de~Almeida \inst{10}\thanks{supported by CAPES Foundation, Ministry of Education of Brazil}
\and A.R.~Bazer-Bachi \inst{11}
\and Y.~Becherini \inst{12,13}
% ...
}
}

\institute{
Universit\"at Hamburg, Institut f\"ur Experimentalphysik, Luruper Chaussee 149, D 22761 Hamburg, Germany \and
Laboratoire de Physique Th\'eorique et Astroparticules, Universit\'e Montpellier 2, CNRS/IN2P3, CC 70, Place Eug\`ene Bataillon, F-34095 Montpellier Cedex 5, France \and
Max-Planck-Institut f\"ur Kernphysik, P.O. Box 103980, D 69029 Heidelberg, Germany \and
Dublin Institute for Advanced Studies, 31 Fitzwilliam Place, Dublin 2, Ireland \and
National Academy of Sciences of the Republic of Armenia, Yerevan  \and
Yerevan Physics Institute, 2 Alikhanian Brothers St., 375036 Yerevan, Armenia \and
Universit\"at Erlangen-N\"urnberg, Physikalisches Institut, Erwin-Rommel-Str. 1, D 91058 Erlangen, Germany \and
Nicolaus Copernicus Astronomical Center, ul. Bartycka 18, 00-716 Warsaw, Poland \and
CEA Saclay, DSM/IRFU, F-91191 Gif-Sur-Yvette Cedex, France \and
University of Durham, Department of Physics, South Road, Durham DH1 3LE, U.K. \and
Centre d'Etude Spatiale des Rayonnements, CNRS/UPS, 9 av. du Colonel Roche, BP 4346, F-31029 Toulouse Cedex 4, France \and
Astroparticule et Cosmologie (APC), CNRS, Universit\'{e} Paris 7 Denis Diderot, 10, rue Alice Domon et L\'{e}onie Duquet, F-75205 Paris Cedex 13, France \thanks{(UMR 7164: CNRS, Universit\'e Paris VII, CEA, Observatoire de Paris)} \and
Laboratoire Leprince-Ringuet, Ecole Polytechnique, CNRS/IN2P3, F-91128 Palaiseau, France
% ...
}

\offprints{S.~Kaufmann\\
\email{\href{mailto:S.Kaufmann@lsw.uni-heidelberg.de}{S.Kaufmann@lsw.uni-heidelberg.de}}\\
\email{\href{mailto:fortin@llr.in2p3.fr}{fortin@llr.in2p3.fr}} \\
\email{\href{mailto:wmcconvi@umd.edu}{wmcconvi@umd.edu}}}

   \date{Received 19 November 2010; accepted 15 March 2011}

\abstract{The high-frequency peaked BL Lac object PKS~2005-489 was the target of a multi-wavelength campaign with simultaneous observations in the TeV $\gamma$-ray (H.E.S.S.), GeV $\gamma$-ray ({\it Fermi}/LAT), X-ray ({\it RXTE}, {\it Swift}), UV ({\it Swift}) and optical (ATOM, {\it Swift}) bands. This campaign was carried out during a high flux state in the synchrotron regime. The flux in the optical and X-ray bands reached the level of the historical maxima. The hard GeV spectrum observed with {\it Fermi}/LAT connects well to the very high energy (VHE, E$>100\;\rm{GeV}$) spectrum measured with H.E.S.S. with a peak energy between $\sim 5$ and 500 GeV.
Compared to observations with contemporaneous coverage in the VHE and X-ray bands in 2004, the X-ray flux was $\sim 50$ times higher during the 2009 campaign while the TeV $\gamma$-ray flux shows marginal variation over the years. 
The spectral energy distribution during this multi-wavelength campaign was fit by a one zone synchrotron self-Compton model with a well determined cutoff in X-rays. 
The parameters of a one zone SSC model are inconsistent with variability time scales.
The variability behaviour over years with the large changes in synchrotron emission and small changes in the inverse Compton emission does not warrant an interpretation within a one-zone SSC model despite an apparently satisfying fit to the broadband data in 2009.
}

   \authorrunning{HESS Collaboration}

   \maketitle
%

\section*{Abstract}
The high-frequency peaked BL Lac object PKS~2005-489 was the target of a multi-wavelength campaign with simultaneous observations in the TeV $\gamma$-ray (H.E.S.S.), GeV $\gamma$-ray ({\it Fermi}/LAT), X-ray ({\it RXTE}, {\it Swift}), UV ({\it Swift}) and optical (ATOM, {\it Swift}) bands. This campaign was carried out during a high flux state in the synchrotron regime. The flux in the optical and X-ray bands reached the level of the historical maxima. The hard GeV spectrum observed with {\it Fermi}/LAT connects well to the very high energy (VHE, E$>100\;\rm{GeV}$) spectrum measured with H.E.S.S. with a peak energy between $\sim 5$ and 500 GeV.
Compared to observations with contemporaneous coverage in the VHE and X-ray bands in 2004, the X-ray flux was $\sim 50$ times higher during the 2009 campaign while the TeV $\gamma$-ray flux shows marginal variation over the years. 
The spectral energy distribution during this multi-wavelength campaign was fit by a one zone synchrotron self-Compton model with a well determined cutoff in X-rays. 
The parameters of a one zone SSC model are inconsistent with variability time scales.
The variability behaviour over years with the large changes in synchrotron emission and small changes in the inverse Compton emission does not warrant an interpretation within a one-zone SSC model despite an apparently satisfying fit to the broadband data in 2009.

\section{Introduction}
PKS~2005-489 is one of the brightest high-frequency peaked BL Lac objects (HBL) in the southern hemisphere. It is located at $\alpha_{\rm{J2000}} = 20^{\rm{h}} 09^{\rm{m}} 25.39^{\rm{s}}$, $\delta_{\rm{J2000}} = -48^\circ 49' 53.7''$ \citep{Johnston1995} and has a redshift of $z=0.071$ \citep{Falomo1987}. 

HBL are characterized by two peaks in their spectral energy distribution (SED) which are located in the UV-X-ray and the GeV-TeV band, respectively. These are commonly explained by leptonic models (e.g. \citealt{Marscher1985}) as synchrotron and inverse Compton (IC) emission from a population of relativistic electrons upscattering their self-produced synchrotron photons (Synchrotron Self Compton models (SSC)).
Also alternative models based on hadronic interactions exist, e.g. \cite{Mannheim1993}.

PKS~2005-489 was detected through the Parkes 2.7~GHz survey \citep{Wall1975} and is part of the 1~Jy catalog \citep{Kuehr1981} of the brightest extragalactic radio sources. 

It has been observed during several years by different X-ray satellites and showed very large flux variations in combination with distinct spectral changes. In October-November 1998, a large X-ray flare  was detected and monitored with {\it RXTE} \citep{Perlman1999}. Shortly before this flare occured, {\it BeppoSAX} observations were conducted which revealed a curved X-ray spectrum from $0.1$ to $200\;\rm{keV}$ with photon indices of $\Gamma_1=2$, $\Gamma_2=2.2$ and a break around $2~\rm{keV}$ \citep{Tagliaferri2001}.

The first evidence for $\gamma$-ray emission was marginally detected ($4.3\sigma$) with the EGRET instrument revealing a flux of $F(>100\;\rm{MeV}) = (1.3 \pm 0.5) \times 10^{-7} \; \rm{cm^{-2}\;s^{-1}}$ \citep{Lin1999,Nandikotkur2007}. This made PKS~2005-489 one of the few HBL detected by EGRET.
Very high energy (VHE, E $>100\;\rm{GeV}$) $\gamma$-rays from PKS~2005-489 were first detected by the Cherenkov telescope array H.E.S.S. (High Energy Stereoscopic System) \citep{Aharonian2005}.
Multi-year studies of the TeV emission by H.E.S.S. together with several multi-wavelength observations are described by the \cite{Aharonian2009}. The VHE $\gamma$-ray spectra can be described by power laws with photon indices varying between 2.9 and 3.7 and hence they are amongst the softest spectra of all VHE $\gamma$-ray active galactic nuclei (AGN). 

Together with the {\it Fermi} Gamma-ray Space Telescope (in operation since June 2008) 
it is possible to determine 
the inverse Compton emission peak of PKS~2005-489. Together with simultaneous broadband observations, the underlying emission processes can be studied in more detail.
In this paper the results of such a broadband simultaneous multi-wavelength campaign on PKS~2005-489 conducted in 2009 are presented.

\section{Multi-wavelength observations and data analysis}

\begin{figure}[h]
\centering
\includegraphics[width=\columnwidth]{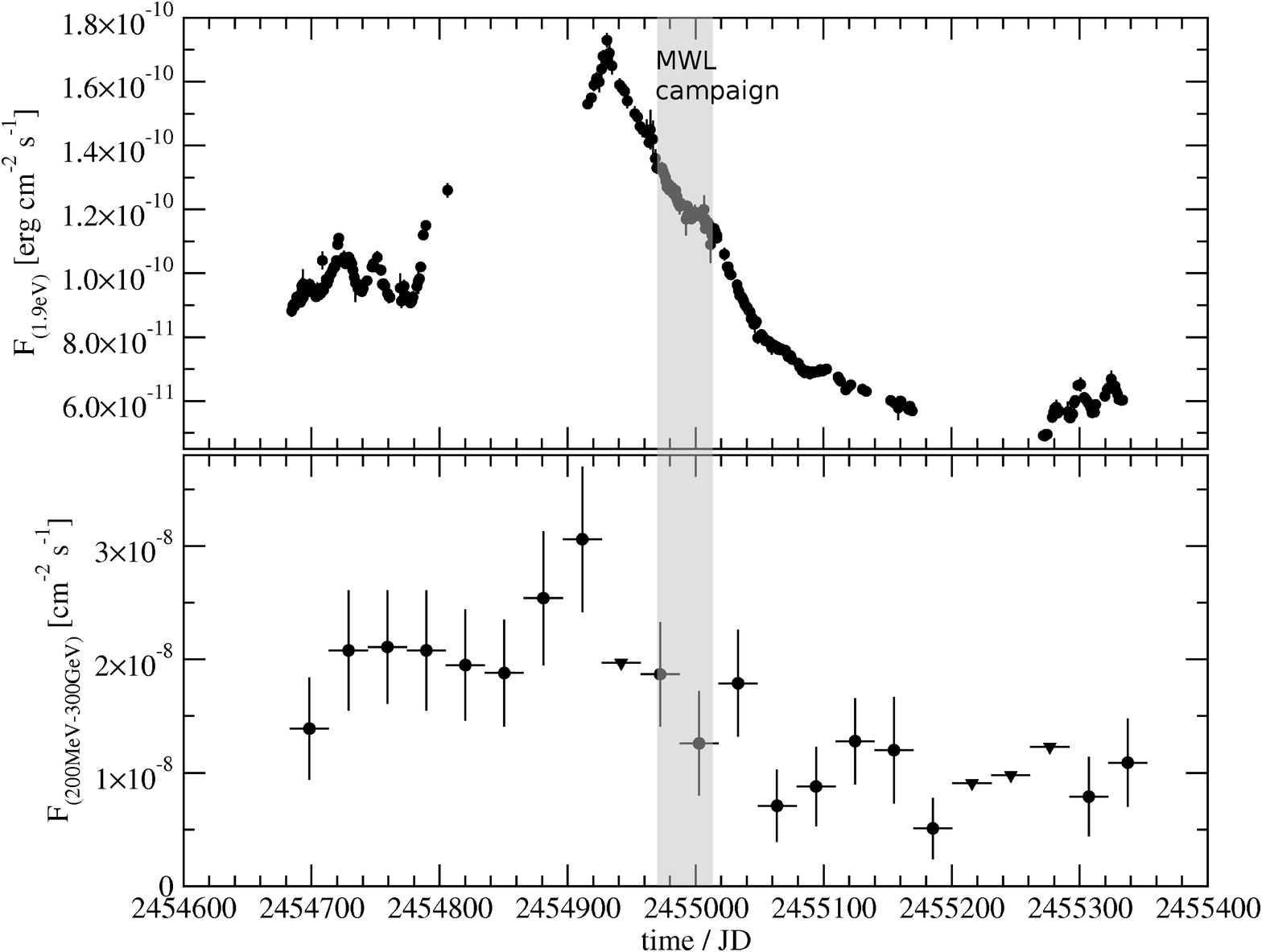}
\caption{Long term lightcurves of PKS~2005-489 over 22 months of the optical emission (upper panel) by ATOM and the HE $\gamma$-ray emission (lower panel) by {\it Fermi}/LAT. The HE $\gamma$-ray lightcurve is binned with 30 days interval. The triangles represent upper limits. The major gaps in the optical lightcurve are due to solar conjunction. The grey band indicates the time of the multi-wavelength campaign from May 22 to July 2, 2009.}
\label{long_LC}
\end{figure}

A multi-wavelength campaign on PKS~2005-489 was conducted from May 22 to July 2, 2009 with observations by the Cherenkov telescope array H.E.S.S. (High Energy Stereoscopic System), the X-ray satellites {\it RXTE} (Rossi X-ray Timing Explorer) and {\it Swift} and the optical 75-cm telescope ATOM (Automatic Telescope for Optical Monitoring for H.E.S.S.). The LAT (Large Area Telescope) instrument onboard the {\it Fermi} Gamma-ray Space Telescope scans the whole sky within approximately 3 hours and hence PKS~2005-489 was regularly monitored during this campaign such that simultaneous information about the brightness and the spectrum of the high energy (HE, $100\;\rm{MeV}<\rm{E}<100\;\rm{GeV}$) $\gamma$-ray emission could be obtained. 
Hence, for the first time, simultaneous observations have been taken on PKS~2005-489 in the VHE, HE $\gamma$-ray, X-ray, UV and optical bands, that can be used for variability and spectral studies.
The simultaneous monitoring by Fermi in the GeV and by ATOM in the optical band over a time of 22 months allows the study of the long term behaviour of PKS~2005-489. The time of the multi-wavelength campaign is marked in the long term light curve shown in Fig.~\ref{long_LC}. 

\begin{figure}[h]
\centering
\includegraphics[width=\columnwidth]{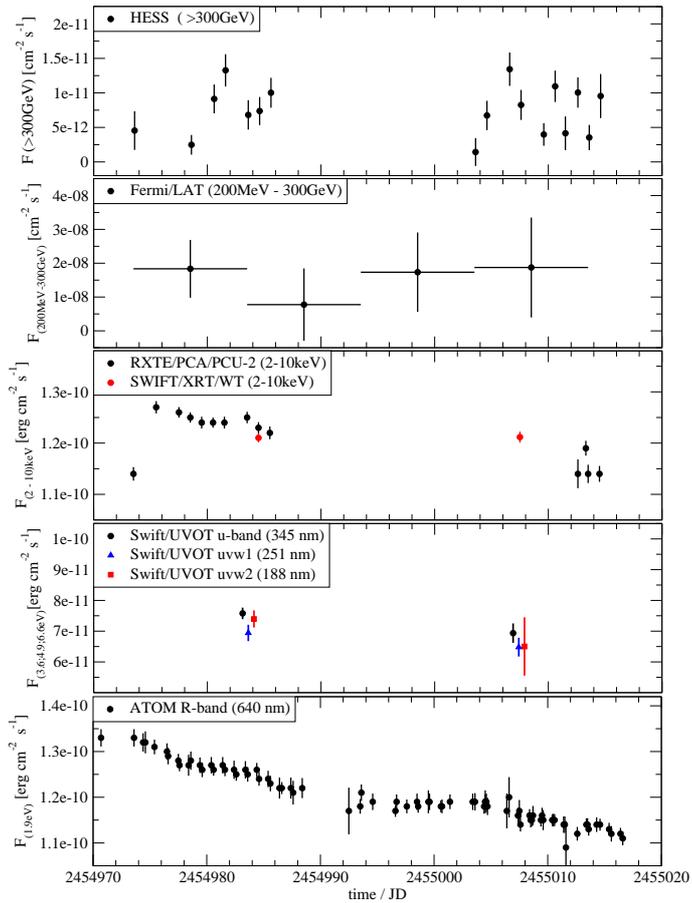}
\caption{Lightcurves of PKS~2005-489 during the multi-wavelength campaign from May 22 to July 2,  2009 with observations by the Cherenkov telescope array H.E.S.S., the Gamma-ray space telescope {\it Fermi}, the X-ray satellites {\it RXTE} and {\it Swift} and the 75-cm optical telescope ATOM. For a better display, the u and uvw2 band data by {\it Swift} are shifted by $\pm 0.5$ days. For the H.E.S.S., {\it RXTE} and {\it Swift} lightcurves a nightly binning was used and the {\it Fermi}/LAT lightcurve is shown in a 10-day binning.}
\label{MWL_LC}
\end{figure}

\subsection{Very high energy $\gamma$-ray data from H.E.S.S.}

The H.E.S.S. experiment consists of four imaging atmospheric Cherenkov telescopes, located in the Khomas Highland, Namibia ($23^\circ 16' 18''$ South, $16^\circ 30' 00''$ East), at an elevation of $1800\;\rm{m}$ \citep{Aharonian2006}. The observations on PKS~2005-489 have been taken with a mean zenith angle of $27^\circ$ from May 22 to July 2, 2009 with a break around full moon since H.E.S.S. does not observe during moontime. The data have been calibrated as described in \cite{Aharonian2004} and analyzed using the {\it standard cuts} resulting in an energy threshold of $\approx 400\;\rm{GeV}$ and following the method described in \cite{Aharonian2006}. After the standard quality selection 13 h of live time remain. A total of $N_{\rm{ON}}=953$  on-source and $N_{\rm{OFF}}=5513$ off-source events with an on-off normalization factor of $\alpha = 0.0932$ were measured. An excess of $N_\gamma = N_{\rm{ON}}-\alpha N_{\rm{OFF}} = 439$ $\gamma$-rays, corresponding to
a significance of $16\sigma$ (following the method of \citealt{LiMa}), results for PKS~2005-489 within the whole observing period. The {\it Reflected-Region} method \citep{Aharonian2006} was used for background subtraction of the spectrum. 
The same data have been analyzed using a different calibration, resulting in compatible results.

\subsection{High energy $\gamma$-ray data from {\it Fermi}/LAT}

The {\it Fermi} Large Area Telescope (LAT) is a pair-conversion gamma-ray detector
sensitive to photons in the energy range from below 20 MeV to more than 300 GeV
\citep{Atwood2009}. Launched by NASA on June 11, 2008, the LAT began nominal
science operations on August 4, 2008. The LAT observations presented here comprise
all the data taken between August 4, 2008 and  June 4, 2010 (22 months), which fully
covers the H.E.S.S. observations taken from May 22 to July 2, 2009. The LAT Science
Tools\footnote{http://Fermi.gsfc.nasa.gov/ssc/data/analysis/scitools/overview.html
} version v9r15 were utilized with the post-launch instrument response
functions (IRF) P6\_V3\_DIFFUSE. Events with high probability of being photons, those
from the {\it diffuse} class, and with zenith angles $<~105^\circ$ were selected. Time intervals
during which the rocking angle (i.e. the angle {\it Fermi} points north or south of the zenith on alternate orbits during sky survey operations) was larger than $52^\circ$ were excluded 
to avoid contamination from the Earth's limb.
A cut at 200 MeV was used to avoid the larger systematic uncertainties in the analysis at lower energies.
Events with energy between 200 MeV and 300 GeV, and
within a $10^\circ$ region of interest (ROI) centered on the coordinates of PKS~2005-489
 were analyzed with an unbinned maximum likelihood
method \citep{Cash1979,Mattox1996}. 
Some sources from the 1FGL catalog \citep{Abdo2010}  were located outside the ROI but close and bright enough to have a significant impact on the analysis. These sources have been taken into account in the analysis.
The background emission was modeled using
standard Galactic and isotropic diffuse emission models\footnote{http://Fermi.gsfc.nasa.gov/ssc/data/access/lat/\\BackgroundModels.html}. The 1FGL catalog \citep{Abdo2010} and a Test Statistic (TS, where $TS = - 2 \Delta \log(\rm{likelihood})$ between models with and without the source) map of the region around PKS~2005-489 were used to identify all the point sources within the ROI. The point sources and PKS~2005-489 were modeled using a power law of the 
form $F(E) = N_0(E/E_0)^{-\Gamma}$.

The likelihood analysis reveals a point source with a high statistical significance
($\sigma \propto \sqrt{\rm{TS}} > 20$). The best-fit position of this point source was
calculated with {\tt gtfindsrc} ($\alpha_{\rm{J2000}} = 20^{\rm{h}} 09^{\rm{m}} 25.0^{\rm{s}}$, $\delta_{\rm{J2000}} = -48^\circ 49' 44.4''$) and has a $95\%$ containment
radius of $1.5'$, consistent with the coordinates of PKS~2005-489. The highest energy
photon associated with the source has an energy of $\sim 180\;\rm{GeV}$ and was detected on March 10, 2009 (i.e. before the start of the H.E.S.S. observations). The highest energy
photon detected by {\it Fermi}/LAT during the H.E.S.S. multi-wavelength campaign has an energy of $\sim 50\;\rm{GeV}$ and was detected on June 1, 2009.

\subsection{X-ray data from {\it RXTE} and {\it Swift}/XRT}

X-ray observations with the Proportional Counter Array (PCA) detector onboard {\it RXTE} \citep{Bradt1993} were obtained in the energy range $2-60\;\rm{keV}$ from May 22 to June 3 with exposures of $2-4\;\rm{ks}$ per pointing, strictly simultaneous with good quality H.E.S.S. observations in 6 nights.
Due to the high state of PKS~2005-489 during this campaign, additional Target of Opportunity (ToO) observations have been taken with the X-ray satellites {\it Swift} and {\it RXTE}. The XRT detector \citep{Burrows2005} onboard {\it Swift} observed in photon-counting (PC) and windowed-timing (WT) mode in the energy range $0.2-10\;\rm{keV}$ on June 1 and June 24 with $\sim 3\;\rm{ks}$ each. The {\it RXTE} ToO observations were performed from June 30 to July 3, 2009.

For the data analysis of the data obtained with {\it RXTE} and {\it Swift}, the software package HEASoft\footnote{http://heasarc.gsfc.nasa.gov/docs/software/lheasoft/} was used.

Only {\it RXTE}/PCA data of PCU2 and the top layer 1 were taken into account to obtain the best signal-to-noise ratio. The data were filtered to account for the influence of the South Atlantic Anomaly, tracking offsets, and electron contamination using the standard criteria recommended by the {\it RXTE} Guest Observer Facility (GOF). For the count rate of $\sim 9\;\rm{cts/s}$ for this observations, the faint background model provided by the {\it RXTE} GOF for count rates $<40\;\rm{cts/s}$ was used to generate the background spectrum with {\tt pcabackest} and the response matrices were created with {\tt pcarsp}.

The second instrument onboard {\it RXTE}, the HEXTE (High Energy X-ray Timing Experiment) onboard {\it RXTE} is measuring in the energy range 15 to 250 keV. For PKS~2005-489, a signal was detected only in the sum of all observations during the campaign. 
The fit of a power law to the combined PCA and HEXTE spectrum gives the same result as for the PCA spectrum alone. The combined fit is dominated by the more sensitive PCA detector with a better signal to noise ratio. Hence, the HEXTE spectrum is not discussed further.

For the Swift spectral analysis, XRT exposure maps were generated with {\tt xrtpipeline} to account for some bad CCD columns that are masked out on-board. The masked hot columns appeared when the XRT CCD was hit by a micrometeoroid.
Spectra of the {\it Swift} data in PC-mode have been extracted with {\tt xselect} from an annulus region with an outer radius of $0.8'$ at the position of PKS~2005-489, which contains $90\%$ of the PSF at 1.5~keV and an inner radius of $\sim 0.1'$ to avoid pileup. An appropriate background was extracted from a circular region  with radius of $3'$ nearby the source. For the WT-mode, appropriate boxes ($\sim 1.6'\times 0.3'$) covering the region with source photons and a background region with similar size were used to extract the spectra. 
The auxiliary response files were created with {\tt xrtmkarf} and the response matrices were taken from the {\it Swift} package of the calibration database {\tt caldb 4.1.3}\footnote{http://heasarc.gsfc.nasa.gov/docs/heasarc/caldb/caldb\_intro.html}.

A power law of the form $F(E) = N_0(E/E_0)^{-\Gamma}$ was used to fit the X-ray spectra of each pointing obtained by {\it RXTE} in the energy range $3-20 \;\rm{keV}$ resulting in consistent parameters and an average photon index of $\Gamma = 2.46 \pm 0.03$.
During this campaign, no significant change in spectral shape was found.
The detailed spectral analysis and results are discussed in secion \ref{spec_sy}.

\subsection{UV data from {\it Swift}/UVOT}

The UVOT instrument \citep{Roming2005} onboard {\it Swift} measures the UV emission in the bands u (345 nm), uvw1 (251 nm) and uvw2 (188 nm) simultaneous to the X-ray telescope with an exposure of $\sim 1\;\rm{ks}$ each. The instrumental magnitudes and the corresponding flux (conversion factors see \citealt{Poole2008}) are calculated with {\tt uvotmaghist} taking into account all photons from a circular region with radius $5''$ (standard aperture for all filters). An appropriate background was determined from a circular region with radius $40''$ near the source region without contamination of sources.

The measured UV fluxes have been corrected for dust absorption using E(B-V)$=$0.056 mag \citep{Schlegel1998} and the $A_\lambda / E(B-V)$ ratios given in \cite{Giommi2006} resulting in a correction of $35\%,29\%,24\%$ for uvw2, uvw1 and u-band, respectively. 
The contribution of the host galaxy to the measured flux is small compared to the correction for extinction and was not taken into account.

\subsection{Optical data from ATOM}

The 75-cm telescope ATOM \citep{Hauser2004}, located at the H.E.S.S. site in Namibia, monitored the flux in the 4 different filters: B (440 nm), V (550 nm), R (640 nm) and I (790 nm) according to \cite{Bessel1990}. The obtained data have been analyzed using an aperture of $4''$ radius and differential photometry with three nearby reference stars from the USNO catalog \citep{Monet1998} 
to determine the apparent magnitudes. 

The host galaxy of PKS~2005-489 is a giant elliptical galaxy with a brightness of $14.5\;\rm{mag}$ (R-band) and a half-light radius of $5.6''$ \citep{Scarpa2000}. In order to correct for the host galaxy light, a de Vaucouleur profile of the galaxy was assumed and rescaled to the aperture used in the ATOM photometry. To calculate the influence of the host galaxy in the B,V and I filter, the spectral template for a nearby elliptical galaxy by \cite{Fukugita1995} has been used. 
The galactic extinction calculated for the ATOM filters is negligible compared to the host galaxy contribution.

\section{Temporal analysis}
\label{var_tev}

The multiwavelength observations taken from May 22 to July 2 2009 were used to search for variability during the phase of high flux. The lightcurves of this campaign are shown in Fig.~\ref{MWL_LC}.

\subsection{Variation in different energy bands}

Very high energy emission has been detected with a mean flux of $F(>\rm{300GeV}) = (6.7 \pm 0.5) \times 10^{-12}\;\rm{cm^{-2}\; s^{-1}}$ for the time of the campaign. The measured flux level is $\sim 2$ times brighter than during the detection of this source in 2004 by H.E.S.S. \citep{Aharonian2009}. 
While no significant long-term trend is detected during the six weeks of observations, a constant flux provides a poor fit with a $\chi^2$/dof=50/16 and a probability of $p<0.1\%$. The nightly binned flux shows evidence for variability of about a factor of 2. The low signal to noise ratio does not allow a precise determination of time scales.

 PKS~2005-489 was 
detected by the LAT during the period corresponding to the H.E.S.S. multi-wavelength
campaign (TS = 69). 
Due to the rather faint HE emission, the binning of the light
curve for the time period of this campaign was chosen to be 10 days. Within the limited statistics, no variations were detected on these short time scales.

The high X-ray flux of PKS~2005-489 is comparable to the historical maximum of 1998 \citep{Perlman1999,Tagliaferri2001}. 
The X-ray flux increased by $ 10\%$ within the first 2 days of the campaign and decreased until the end of the campaign back to the initial flux level.

The nightly binned u, uvw1 and uvw2-band observations with the {\it Swift}/UVOT detector do not show any variation.

The high optical flux measured by ATOM in all 4 filters decreases monotonically by $\sim20\%$ during this campaign. The colors remained constant with averages of B-R$=1.36$ mag, V-R$=0.78$ mag and R-I$=0.08$ mag.

The long term lightcurves of the optical and HE $\gamma$-ray emission of PKS~2005-489 over 22 months is shown in Fig.~\ref{long_LC}. The maximum in the optical band is clearly identified at 2454930.7 JD. The {\it Fermi}/LAT light curve displays a variation in the monthly binning with an amplitude similar to the optical variation.
A fit with a constant flux results in a poor description with a $\chi^2$/dof =  41 / 17 and a probability of $p<1\%$. The decrease in the second half of the light curve has a significance of $\sim 4.5\sigma$. In the monthly binning of the light curve, the identification of the maximum is uncertain by $\sim 30$ day. Considering this, it is marginally consistent with the maximum of the synchrotron emission.

\subsection{Correlations}

The highest flux measured in the X-ray band seems to follow the high flux measured in the optical before the beginning of this campaign. In both wavebands a slow decrease could be measured over the time period of this campaign, where the optical flux decrease by $\sim 20\%$ and the X-ray flux by $\sim 10\%$. The correlation coefficient (also known as Pearson's correlation) of the optical R-band from ATOM and the X-ray flux from {\it RXTE} is 0.7 for the full time range. Considering only the decaying part of the X-ray lightcurve, this factor increases to 0.96.
A change of similar amplitude could not be detected with the {\it Fermi}/LAT due to the large uncertainties.

Interestingly, the variation in the VHE flux, e.g. the increase in flux between 2454978 and 2454982 JD is not seen in the simultaneous X-ray and optical observations which do not vary significantly during this period.

\section{Spectral energy distribution}

\begin{figure}[t]
\centering
\includegraphics[width=\columnwidth]{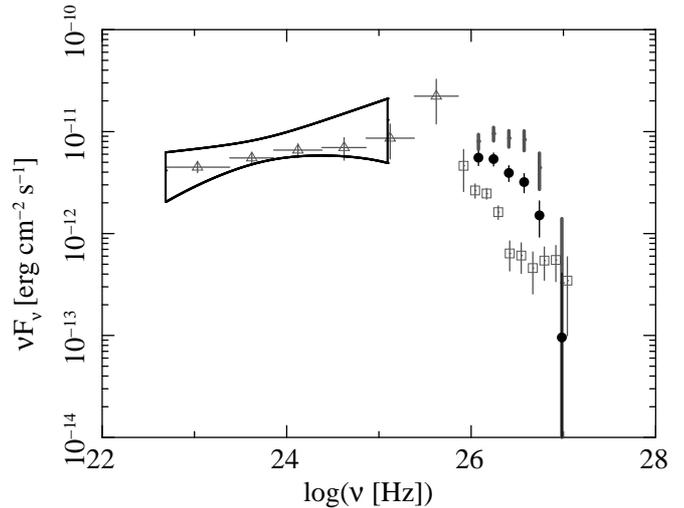}
\caption{The $\gamma$-ray energy spectra of PKS~2005-489 covering the inverse Compton peak of the spectral energy distribution. In  black, the $\gamma$-ray spectra ({\it Fermi}/LAT as butterfly and H.E.S.S. as circles) of this simultaneous multi-wavelength campaign are denoted. In grey open symbols, historical and time-averaged data are shown: triangles represent the {\it Fermi}/LAT spectrum from 22 months of observations and the squares show the integrated HESS spectrum extracted from observations from 2004 to 2007 \citep{Aharonian2009}. 
Dark grey bars denote the VHE $\gamma$-ray spectrum of 2009, corrected for the absorption by the extragalactic background light. The size of the bar reflects the highest and smallest correction, including the statistical and systematic uncertainties, using the models by \cite{Aharonian2006EBL} and \cite{Franceschini2008}.
} 
\label{SED_IC}
\end{figure}

For the first time, simultaneous observations from optical to VHE have been obtained on PKS~2005-489,
providing a very good coverage of the emission peaks seen in the spectral energy distribution (SED) 
(see Fig.~\ref{SED_IC} and Fig.~\ref{SED_SSC}).

\subsection{Spectral data in the $\gamma$-ray range}

The measured very high energy spectrum by H.E.S.S. (see Fig.~\ref{SED_IC}) during the period of this campaign, can be described by a power law ($N(E) = N_0 \times (E/E_0)^{-\Gamma}$) with a normalization of $N_0 = (2.1 \pm 0.2_{\rm{stat}} \pm 0.4_{\rm{sys}} ) \times 10^{-12}\;\rm{cm^{-2}\; s^{-1} \; TeV^{-1}}$ at $E_0 = 1\;\rm{TeV}$ and a spectral index of $\Gamma = 3.0 \pm 0.1_{\rm{stat}} \pm 0.2_{\rm{sys}}$ ($\chi^2 /$dof = 13/4 ).
The subscripts refer to statistical
and systematic uncertainties.  
A power law with an exponential cutoff with a normalization of $N_0 = (1.1 \pm 0.6_{\rm{stat}} \pm 0.2_{\rm{sys}}) \times 10^{-11}\;\rm{cm^{-2}\; s^{-1} \; TeV^{-1}}$ at $E_0=0.5\;\rm{TeV}$, a photon index of $\Gamma = 1.3 \pm 0.6_{\rm{stat}} \pm 0.2_{\rm{sys}}$ and a cutoff at $E_{\rm{cutoff}}=1.3\pm 0.5_{\rm{stat}}\;\rm{TeV}$ is a much better description than the simple power law.

This VHE spectrum has been corrected for the absorption by the extragalactic background light using the models by \cite{Aharonian2006EBL} and \cite{Franceschini2008}. In Fig.~\ref{SED_IC} the minima and maxima of this correction are shown to illustrate the uncertainties of this correction. Since the measurement errors cover the uncertainties using different EBL models, only one model has been chosen for the overall SED shown in Fig.~\ref{SED_SSC}.

The time averaged very high energy spectrum from 2004-2007  (shown in Fig.~\ref{SED_IC}) can be described by a power law with a spectral index of $\Gamma =  3.2 \pm 0.16_{\rm{stat}} \pm 0.1_{\rm{sys}}$, consistent with the one obtained during this campaign but does not show significant curvature \citep{Aharonian2009}. 

During this campaign PKS~2005-489 shows a marginally harder TeV spectrum than during its TeV detection in 2004 \citep{Aharonian2009} when the softest spectrum of a TeV blazar with spectral index of $\Gamma = 3.7 \pm 0.4_{\rm{stat}} \pm 0.1_{\rm{sys}}$ was measured. This spectrum, obtained in 2004, with contemporaneous X-ray, UV and optical observations is shown in Fig.~\ref{SED_SSC} for comparison.
Most noticeable in the comparison is the indication of a cutoff in the VHE spectrum of this campaign and the different normalizations which suggest a change of the inverse Compton peak to higher energies in 2009.
The VHE spectrum resulting from this multi-wavelength campaign and from 2004 have been corrected for the absorption by the extragalactic background light (EBL) using the model by \cite{Franceschini2008} and are shown in Fig.~\ref{SED_SSC}.

The averaged {\it Fermi}/LAT photon spectrum during the time of this campaign is fitted by a power law for which the
normalization constant $N_0$ is $(0.62 \pm 0.04_{\rm{stat}} \pm 0.02_{\rm{sys}})\times 10^{-12} \rm{cm^{-2} \; s^{-1} \; MeV^{-1}}$, the spectral index $\Gamma$ is $1.79 \pm 0.05_{\rm{stat}} \pm 0.07_{\rm{sys}}$, and $E_0 = 2385 \;\rm{MeV}$ is the
energy at which the correlation between the fitted values of the normalization
constant and the spectral index is minimized. Changing the model of the spectrum to a log-parabola
does not improve the quality of the spectral fit significantly. Spectral points were
obtained by dividing the data into  six equal logarithmically-spaced energy bands. A
separate likelihood analysis was run over each band. For all sources in the model of the region, the index was frozen to the time-independent best-fit value, and the flux was left free to vary.

PKS~2005-489 is too faint in the HE band to obtain spectral points for the epoch of
the H.E.S.S. multi-wavelength campaign. A 1-sigma error contour (butterfly) was
calculated using the covariance matrix produced during the {\tt gtlike} likelihood fit
\citep{Abdo2009}. The full energy range (200 MeV - 300 GeV) was used for the fit
but the butterfly was extended to only 50 GeV to include the highest energy photon
detected during the multi-wavelength campaign.

The HE spectral points obtained for the  22-month period and the butterfly calculated
for the H.E.S.S. multi-wavelength campaign are shown in Fig.~\ref{SED_IC}.

\subsection{Spectral data in the synchrotron range}
\label{spec_sy}

\begin{figure}[t]
\centering
\includegraphics[width=\columnwidth]{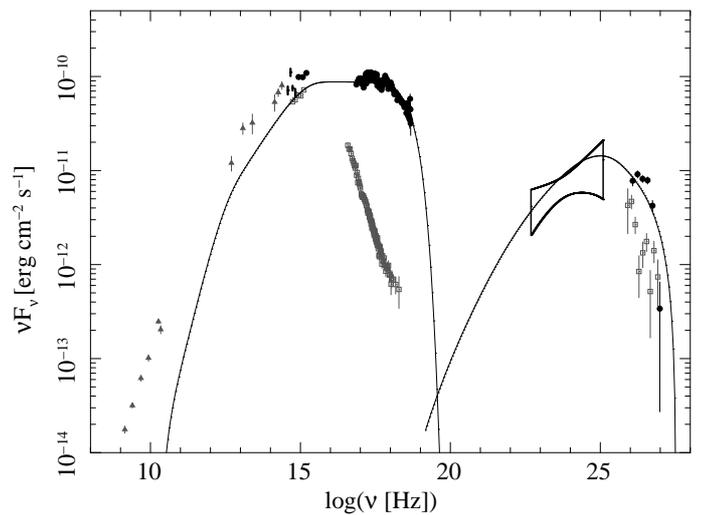}
\caption{
Spectral energy distribution of PKS2005-489 during this multi-wavelength campaign with simultaneous observations by HESS, {\it Fermi}/LAT, {\it RXTE}, {\it Swift} and ATOM (black symbols) using the time range shown in Fig. 1. The spectra shown here are corrected for Galactic extinction, $N_H$ absorption, and EBL absorption. The size of the symbols for the optical data represent the flux range measured. The deabsorbed X-ray and TeV spectra of PKS~2005-489 in 2004 \citep{Aharonian2009}, as well as historical IR and radio observations are shown in grey.
The black curves represent a one zone SSC model as described in the text.
} 
\label{SED_SSC}
\end{figure}

PKS~2005-489 was in a very high X-ray state during this campaign with a maximum flux comparable to the historical maximum of 1998. 
The spectrum during the 2009 campaign was determined from the sum of all {\it RXTE} and {\it Swift} observations. In the energy range of $0.3 - 20\;\rm{keV}$ it can be described by a broken power law with $\Gamma_1 = 2.02 \pm 0.01$, $\Gamma_2 = 2.46 \pm 0.01$, break energy at $3.2 \pm 0.2\; \rm{keV}$ and a normalisation of $N_0 = (5.80\pm 0.06)\times 10^{-2} \;\rm{cm^{-2}\; s^{-1} \; keV^{-1}}$ at $E_0 = 1 \;\rm{keV}$ ($\chi^2$/dof = 582/373), using the Galactic absorption of $3.94\times10^{20}\;\rm{cm^{-2}}$ (LAB Survey, \citealt{Kalberla2005}) as a fixed parameter.
This spectrum by {\it RXTE} and {\it Swift} corrected for the Galactic absorption is shown in Fig.~\ref{SED_SSC}.
The integrated flux between 2 and 10 keV is $F_{(2-10\;\rm{keV})} = (1.23 \pm 0.01)\times 10^{-10} \; \rm{erg \; cm^{-2} \; s^{-1}}$ which is a factor of $\sim 100$ higher than the integrated flux of the XMM-Newton observation in 2004 \citep{Aharonian2009} while the monochromatic flux at 2~keV is $\sim 50$ times higher than in 2004.

The two X-ray instruments had different sampling patterns. Since variability was detected during the entire campaign, a joint spectral fit was obtained on June 1, 2009. The derived broken power law model was fit, resulting in $\Gamma_1 = 2.04 \pm 0.02$, $\Gamma_2 = 2.50 \pm 0.07$, break energy at $3.8 \pm 0.4\; \rm{keV}$ and a normalisation of $N_0 = (5.72\pm 0.07)\times 10^{-2} \;\rm{cm^{-2}\; s^{-1} \; keV^{-1}}$ at $E_0 = 1 \;\rm{keV}$ ($\chi^2$/dof = 493/302), which is consistent to the fit of the summed spectra. The integrated flux is $F_{(2-10\;\rm{keV})} = (1.23 \pm 0.01)\times 10^{-10} \; \rm{erg \; cm^{-2} \; s^{-1}}$.

In Fig.~\ref{SED_SSC} a dip at the low energy end of the X-ray spectrum can be seen. As discussed by \cite{Godet2009}, the most likely explanation for this feature is a detector effect.


The spectrum obtained during the historical maximum  on November 10, 1998 is described by a power law of $\Gamma = 2.35 \pm 0.02$ with a flux of $F_{(2-10\;\rm{keV})} = 3.3\times 10^{-10}\;\rm{erg \; cm^{-2} \; s^{-1}}$\citep{Perlman1999}. The {\it BeppoSAX} spectrum, obtained eight days before on November 1-2, could be fitted with a broken power law with $\Gamma_1=2.02 \pm 0.04$, $\Gamma_2=2.21 \pm 0.02$, $E_{\rm{break}}=1.9 \pm 0.4 \;\rm{keV}$ ($F_{(2-10\;\rm{keV})} = 1.8\times 10^{-10}\;\rm{erg \; cm^{-2} \; s^{-1}}$ ) using a fixed Galactic absorption of $4.2\times 10^{20} \;\rm{cm^{-2}}$ \citep{Tagliaferri2001}.
Compared to the spectral shape of these observations, the high energy photon index of the broken power law detected in the 2009 campaign is higher ($\Delta \Gamma \approx 0.3$) and the peak is at a slightly higher energy. Since the spectrum at the time of the historical maximum was obtained with {\it RXTE} in the energy range $2-10\;\rm{keV}$, a spectral break $<2-3\; \rm{keV}$ cannot be ruled out.

IR observations have been obtained between September 28 and October 1, 1998 with the 2.5m Telescope at the Las Campanas Observatory using the NIR camera with $\rm{J_s}$ (1.24~$\mu$m), H (1.65~$\mu$m), $\rm{K_s}$ (2.16~$\mu$m) \citep{Cheung2003}. The observed fluxes were corrected for the influence of the host galaxy in these bands and are shown in Fig.~\ref{SED_SSC}. PKS~2005-489 is also detected in the frequencies $12~\mu$m, $25~\mu$m, $60~\mu$m with the Infrared Astronomical Satellite (IRAS) as mentioned in the IRAS faint source catalog v2\footnote{http://vizier.cfa.harvard.edu/viz-bin/Cat?II/156A} \citep{Moshir1990}.

Radio observations of PKS~2000-489 with the Australian Telescope Compact Array (ATCA) at the frequencies  8.6, 4.8, 2.5, 1.4 GHz have been done from October 1996 to February 2000 \citep{Tingay2003}. It was also observed with ATCA in the frequencies 18.5 GHz and 22 GHz in March 2002 \citep{Ricci2006} during measurements of all sources of the 5GHz 1Jy-catalog \citep{Kuehr1981}. 

PKS~2005-489 is known to be variable in the synchrotron range, therefore the historical IR and radio observations are not taken into account for the SED modelling.

\subsection{SED model}
 \label{section_sed}
The multi-wavelength data obtained during this campaign cover well the two emission peaks in the spectral energy distribution (see Fig.~\ref{SED_SSC}) allowing the simultaneous determination of the peak energies and fluxes of the two spectral components.

A one zone SSC model using the code by \cite{Krawczynski2004} has been applied to create a model that can describe the simultaneously observed multi-wavelength SED of this campaign. 

The specific spectral shape (broken power law) in the X-ray band is a strong restriction for the cutoff in the synchrotron emission and therefore limits the maximum energy of the electron distribution. The optical, UV and X-ray spectra describing the synchrotron emission restrict the parameters of the electron distribution. This distribution can be described by a broken power law with indices $n_1=2$ and $n_2=3$, as well as the minimum $E_{\rm{min}}=7.9\times 10^{8}$ eV ($\gamma=1.5\times10^{3}$), break $E_{\rm{b}}=2\times 10^{10}$ eV ($\gamma=3.9\times10^{4}$) and maximum energy $E_{\rm{max}}=1\times 10^{12}$ eV ($\gamma=2\times10^{6}$). The high energy power law photon index has been chosen such that the flat shape of the SED in the UV to X-ray range is reproduced. The difference in indices of the broken power law follows the standard break due to cooling ($n_2 - n_1 = 1$).

In order to explain the measured inverse Compton emission in view of the large separation to the synchrotron peak, the remaining parameters describing the emission volume were chosen as 15 for the Doppler factor, $\rm{R} = 4\times 10^{17}~\rm{cm}$  for the radius and $0.02\;\rm{G}$ for the magnetic field. The Doppler factor of 15 was found to be the smallest possible to reproduce the measured shape. The resulting SSC model can well describe the measured broadband SED.
However, this leads to $R / \delta$ by a factor of $\approx 3$ larger (corresponding to a variability time scale of approx. 9 days) than the value determined from the variability time scale of a few days detected for the source. 
In order to correct this, a Doppler factor of around 50 would be needed, but this would lead to a lower synchrotron peak frequency and the cutoff in the synchrotron emission detected in the X-ray band can not be described after adjusting the parameters to reduce the large fluxes resulting from the high Doppler factor.

The parameters of the SSC model optimized for the broadband spectra are close to equipartition (kinetic energy density is $\approx 3\times 10^{-5} \;\rm{erg\; cm^{-3}}$, magnetic energy density is $\approx 2\times 10^{-5} \;\rm{erg\; cm^{-3}}$), although this was not a restriction for the model.
 The energy of the emitting electrons contained in the emission region of the jet, that is responsible for the broadband emission up to very high energies, is E$\approx 9\times 10^{48} \; \rm{erg}$.

Despite the overall match, this one component synchrotron model is disfavoured by the variability characteristics, e.g. the variability time scale of some days, described in section~\ref{var_tev}.
In light of the inability of the one-zone SSC model to explain the variability, the use of such a model as a physical representation of the blazar jet should be considered an approximation.

Comparison of the high synchrotron state in this 2009 multi-wavelength campaign to earlier observations would require changes to the SSC model.
Between 2004 and 2009 the X-ray flux changed by a factor of $\sim 50$ and the spectrum changed from a broken power law with photon indices of $\Gamma_1 = 3.1$ and $\Gamma_2 = 2.6$ and a break at $2.5\;\rm{keV}$ \citep{Aharonian2009} to a broken power law with photon indices $\Gamma_1 = 2.0$ and $\Gamma_2 = 2.5$ and with a break at 3.8 keV as can be seen in Fig.~\ref{SED_SSC}. During this time the TeV emission showed small variation in flux with no significant change in slope.
No trivial change in parameters of the one zone SSC model used to represent the 2009 multi-wavelength spectra can describe the steep X-ray spectrum measured in 2004. The specific change in shape of the synchrotron emission peak can be sketched using indices of $n_1=2$ and $n_2=4.8$ for the broken power law of the electron distribution. But this would mean an arbitrary change of $2.8$ of the photon indices lacking any physics basis.
Multi-component models do not have this problem. However, given the wavelength coverage and the limited signal-to-noise in fine time-bins, multi-component models remain ambiguous.

 As shown in \cite{Aharonian2009}, between 2004 and 2005, the X-ray flux increased by a factor of $\sim 16$ without significant increase of the TeV flux.
With the large X-ray flux state detected in 2009, the difference in the X-ray flux to 2004 is much higher (factor of $\sim 50$) with still only marginal increase of the TeV flux by $\sim 2$.

\section{Summary and Conclusions}

A broadband multi-wavelength campaign on PKS~2005-489 with, for the first time, simultaneous observations in the VHE $\gamma$-ray (by H.E.S.S.), HE $\gamma$-ray ({\it Fermi}/LAT), X-ray ({\it RXTE}, {\it Swift}), UV ({\it Swift}) and optical (ATOM, {\it Swift}) band has been conducted between May 22 and July 2,  2009. PKS~2005-489 was observed in a very bright X-ray state, comparable to the historical maximum. The optical flux at the beginning of the campaign was also the highest in several years. For the first time such a high state in synchrotron emission for PKS~2005-489 was covered by $\gamma$-ray observations with {\it Fermi}/LAT and H.E.S.S. Variability in the VHE $\gamma$-ray emission and a decrease in flux in the X-ray and optical bands were detected. 
Considering a longer time range of 22 months, variations could be detected in the HE $\gamma$-ray band with an amplitude similar to the variation of the optical emission.

The simultaneously measured hard HE spectrum connects well to the VHE spectrum resulting in a peak of the inverse Compton emission between $\sim 5$ and $\sim 500$ GeV. These characteristics are compatible with the long term behaviour in the $\gamma$-ray band, while the inverse Compton peak is at higher energies during this campaign due to the higher VHE flux and the curved spectrum measured.
The observed X-ray spectrum obtained during the campaign shows a clear break at $\sim 4$ keV indicating the cutoff of the synchrotron emission. The hard photon index up to the break energy is one of the main differences of previous X-ray observations in a lower flux state.

These multi-wavelength observations cover well the two emission peaks in the spectral energy distribution. A one zone SSC model was used to fit the broadband spectra.
The characteristic broken power law shape of the X-ray spectrum yields a clear break of the synchrotron emission, restricting the parameter range of the SSC model.

In comparison with previous observations in the X-ray and TeV $\gamma$-ray range of 2004 \citep{Aharonian2009}, the X-ray flux has changed by a factor of $\sim 50$ while the VHE $\gamma$-ray flux shows smaller variations by a factor of $\sim 2$. This large change in synchrotron emission with very small changes in inverse Compton emission is unusual for high-frequency peaked BL Lac objects. 
Especially the change in spectral shape in the X-ray range between 2004 and 2009 is not reflected in the inverse Compton range. 

The prominent TeV blazars, Mrk 421 and PKS 2155-304, have been observed simultaneously in the X-ray and VHE band several times. An interesting aspect of study is the relation of the X-ray and VHE flux during different flaring states of these sources. For PKS 2155-304 a cubic relation ($F_\gamma \propto F_{\rm{X}}^{3}$) was found during a flare in 2006 \citep{Aharonian2009b}, which means that the VHE flux showed a much stronger variation than the X-ray flux. For Mrk 421, a period with several flares occured in 2001, which was covered by observations of {\it RXTE}, Whipple and HEGRA \citep{Fossati2008}. The resulting relations of the X-ray and VHE flux changed from $\beta=0.5$ to $\beta=2$ ($F_\gamma \propto F_{\rm{X}}^{\beta}$) \citep{Fossati2008}.
The simultaneous X-ray and VHE measurement of PKS~2005-489 in 2004 and 2009 exhibit a different behaviour. The X-ray flux varies strongly while VHE fluxes hardly change
which yield  $\beta=0.2$. Contrary to the other sources, 
this may imply that during the flare the electron population changes such that the IC scattering occurs predominantly in the Klein-Nishina regime, which has lower efficiency for production of TeV photons, while it is (largely) in the Thomson regime during quiescence.

The huge changes in synchrotron emission with small changes in inverse Compton emission over the years between low and high flux states in the synchrotron branch of PKS~2005-489 challenge the applicability of a one zone SSC model to this source. A similar conclusion about the limitations of a one-zone SSC model was reached for the multi-wavelength campaign on PKS~2155-304 
\citep{Aharonian2009c}. 
Results such as these emphasize the need for sustained or repeated multi-wavelength campaigns to measure spectral variability.

\begin{acknowledgements}
The support of the Namibian authorities and of the University of Namibia in facilitating the construction and operation of HESS is gratefully acknowledged, as is the support by the German Ministry  for Education and Research (BMBF), the Max Planck Society, the French Ministry for Research, the CNRS-IN2P3 and the Astroparticle Interdisciplinary Programme of the CNRS, the U.K. Science and Technology Facilities Council (STFC), the IPNP of the Charles University, the Polish Ministry of Science and Higher Education, the South African Department of Science and Technology and National Research Foundation, and by the University of Namibia. We appreciate the excellent work of the technical support staff in Berlin, Durham, Hamburg, Heidelberg, Palaiseau, Paris, Saclay, and in Namibia in the construction and operation of the equipment. \\
\\
The \textit{{\it Fermi}} LAT Collaboration acknowledges generous ongoing support
from a number of agencies and institutes that have supported both the
development and the operation of the LAT as well as scientific data analysis.
These include the National Aeronautics and Space Administration and the
Department of Energy in the United States, the Commissariat \`a l'Energie Atomique
and the Centre National de la Recherche Scientifique / Institut National de Physique
Nucl\'eaire et de Physique des Particules in France, the Agenzia Spaziale Italiana
and the Istituto Nazionale di Fisica Nucleare in Italy, the Ministry of Education,
Culture, Sports, Science and Technology (MEXT), High Energy Accelerator Research
Organization (KEK) and Japan Aerospace Exploration Agency (JAXA) in Japan, and
the K.~A.~Wallenberg Foundation, the Swedish Research Council and the
Swedish National Space Board in Sweden.

Additional support for science analysis during the operations phase is gratefully
acknowledged from the Istituto Nazionale di Astrofisica in Italy and the Centre
National d'\'Etudes Spatiales in France.\\
\\
The authors acknowledge the support by the {\it RXTE} and {\it Swift} teams for providing ToO observations and the use of the public HEASARC software packages. \\
S.K. and S.W. acknowledge support from the BMBF through grant DLR 50OR0906.
\end{acknowledgements}
%
\bibliographystyle{bibtex/aa}
\bibliography{ref2.bib}

\end{document}